\documentclass[12pt]{iopart}
\usepackage{graphicx}

\newcommand{\beq}{\begin{equation}}
\newcommand{\eeq}{\end{equation}}

\newcommand{\beqa}{\begin{eqnarray}}
\newcommand{\eeqa}{\end{eqnarray}}

\begin{document}

\title[Embryonic Pattern Scaling]{Embryonic Pattern Scaling Achieved by Oppositely Directed Morphogen Gradients}

\author{Peter McHale, Wouter-Jan Rappel and Herbert Levine}
\address{Department of Physics and Center for Theoretical Biological 
Physics, University of California, San Diego, La Jolla, CA 92093-0374, USA}
\date{\today}

\begin{abstract}
Morphogens are proteins, often produced in a localised region, 
whose concentrations 
spatially demarcate regions of differing gene expression in 
developing embryos. 
The boundaries of expression must be set accurately and 
in proportion to the size of the one-dimensional developing field; 
this cannot be accomplished by
a single gradient. Here, we  show how a pair of 
morphogens produced at opposite
ends of a developing field can solve the pattern-scaling problem.
In the most promising scenario, 
the morphogens effectively interact according to the annihilation reaction 
$A+B\rightarrow\emptyset$ and the switch 
occurs according to the absolute concentration of $A$ or $B$.
In this case embryonic markers across the entire developing field
scale approximately with system size; this cannot be achieved with a pair of 
non-interacting gradients that combinatorially 
regulate downstream genes. This scaling occurs 
 in a window of 
developing-field sizes centred at a few times the morphogen decay length.
\end{abstract}


\maketitle

\section{Introduction}

Morphogen gradients play a crucial role in establishing patterns of gene expression during development. These patterns then go on to determine the complex three-dimensional morphology that is needed for organism functionality. Because not all environmental variation can be controlled, gene patterning must be robust to a variety of perturbations, i.e. must compensate for the unpredictable~\cite{Eldar2004}. 

One aspect of this robustness concerns the notion of size scaling \cite{Wieschaus2005}. Typically, gene patterns are established in proportion to the (variable) size of the nascent embryo. A dramatic demonstration of this was made recently in the case of 
{\sl Drosophila}  where the posterior boundary of the {\sl hunchback} gene expression domain was shown to scale (to within $5\%$) with embryo size \cite{Houchmandzadeh2002}. In the standard model of pattern formation in developmental biology cells acquire their positional information by measuring the concentration 
of a morphogen gradient and comparing to some hard-wired set of thresholds \cite{Wolpert1969,Nusslein-Volhard1988,Terry2002}. As the simplest single-source diffusing morphogen gradient with fixed thresholds clearly does not exhibit this type of proportionality, it is clear that more sophisticated dynamics must be responsible for the observed structures \cite{Bisseling2005}. Unfortunately,  little to nothing is known experimentally about how this 
pattern scaling comes about. 

As a first step in deciphering what these more complex processes might entail, we study here the issue of how two morphogen gradients, directed from 
opposite ends of a developing field, may solve the pattern-scaling 
problem \cite{Wolpert1969}. Operationally, opposing gradients may arise 
in developing systems in at least two ways. First mRNA, from which protein
is translated, may be
anchored at opposite ends of the region in question. As an example, 
in the 
{\sl Drosophila} syncytium an anterior-to-posterior gradient 
is established by the localisation 
of {\sl bicoid} mRNA to the anterior, while 
{\sl nanos} mRNA localised at the posterior defines a reciprocal gradient \cite{StJohnston2004}.
Second, protein may be secreted by clusters of cells, one cluster located at opposite ends of the developing field \cite{Barkai2002}. In both cases we will assume that there is some flux of morphogens entering a specific region and assume that there is no production in the bulk of the region. We further assume that the morphogen reaches neighbouring cells by an effective diffusion process thereby creating a gradient \cite{Julicher2005}. Finally, although time-dependent effects in development patterning might be important in  some contexts \cite{Reinitz2004}, we assume here  that a steady-state analysis is sufficient insofar as scaling of patterns with system size is concerned. 

We consider two mechanisms in which a pair of morphogen gradients 
transmits size information to the developmental pattern.  
The first mechanism, which uses the concentrations of both gradients combinatorially, is an alternative to the simple gradient mechanism \cite{Small2005}. In this mechanism there exist overlapping DNA-binding sites of species $A$ and $B$ in the cis-regulatory modules of the target genes. We note that in the {\sl Drosophila} syncyctium some {\sl kr\"{u}ppel} binding sites overlap extensively with {\sl bicoid} sites \cite{Small2005,Levine1992}.
In this scenario one of the morphogens acts as a transcription factor and the
other acts as an effective repressor by occluding the 
binding site of the first. Hence
 the target gene is switched on according to the relative concentrations
of the two species \cite{Terry2005}. In the second mechanism protein $B$
irreversibly inhibits the activity of transcription factor $A$ by 
either directly degrading it or by irreversibly binding to it.  The interaction is described by the annihilation reaction $A+B\rightarrow \emptyset$.
Here we assume that the target gene measures the absolute value of the 
$A$ concentration as in the standard model of developmental patterning; the $B$ gradient serves only to provide size 
information to the $A$ concentration field. 

The goal of this work is to study these two possibilities and see the extent to which they do in fact solve the pattern scaling problem. To this end we measure the range of variables over which scaling is approximately valid. We begin in \S~\ref{singleGradSect} by pointing out that a single gradient in a finite system cannot set markers proportionately in the developing field. In \S~\ref{combinatorialSect} we study the case of two gradients whose binding sites overlap and show that approximate scaling then occurs in a restricted fraction of the developing field typically located midway between the sources. We then turn to the annihilation model of two gradients in \S~\ref{coupledSect} and show that its scaling performance is excellent throughout the developing field. 

\section{Single Gradient}
\label{singleGradSect}

Let $A$ be 
the concentration of the morphogen which in the simplest model obeys
\beq
0 = D_a \partial_x^2 A - \beta_a A
\eeq
at steady state. Here $D_a$ is the diffusion constant of protein $A$ and 
$\beta_a$ is the degradation rate. 
Molecules of $A$ are injected at the left boundary with rate $\Gamma_a$ and are 
confined to the interval $[0,L]$ by a zero-flux boundary condition
\beq
\begin{array}{ll}
-D_a\partial_x A (0) = \Gamma_a; & -D_a\partial_x A (L) = 0 .
\end{array}
\eeq
The obvious solution is 
\beq
A=\frac{\lambda_a \Gamma _a}{D_a} \left[\sinh \left(  \frac{L}{\lambda_a} \right)\right]^{-1}  \cosh \left( \frac{ L-x}{\lambda_a } \right) \  \ \equiv \ A(L) \cosh \left( \frac{ L-x}{\lambda_a } \right) .
\label{obvEq}
\eeq
The length scale $\lambda_a$ is defined by 
$\lambda_a =  \sqrt{D_a/\beta_a}$.

\begin{figure}
\centerline{\includegraphics[scale=0.8]{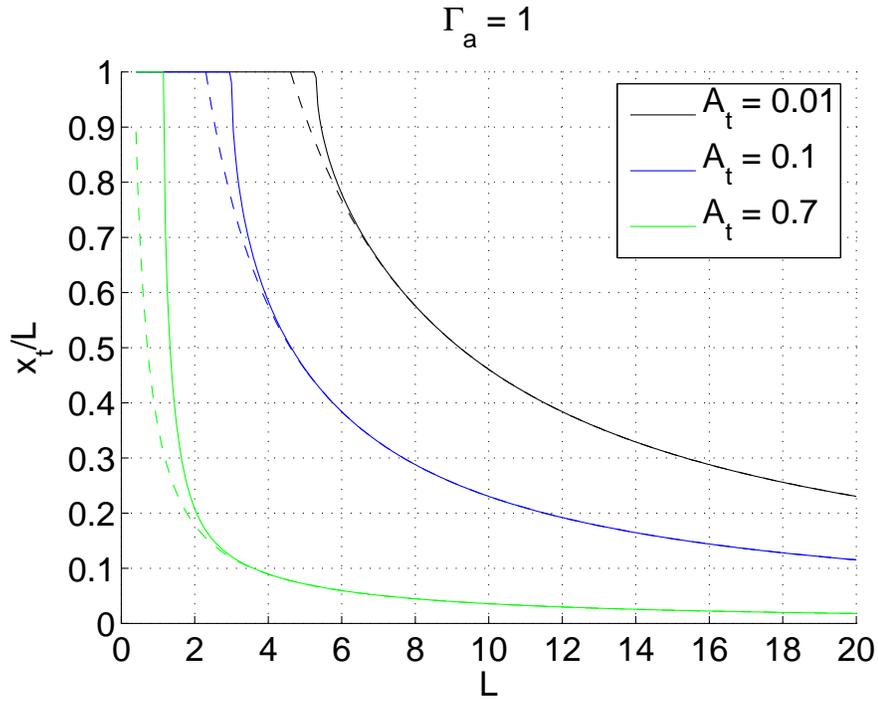}}
\caption{Dependence of normalised $x_t$ on system length $L$
for the case of a single gradient.
The solid lines are the 
analytic expressions Eq.~(\ref{xtLowEq}) for $x_t/L$ for values 
of the threshold concentration equal to (from top to bottom) $A_t = 0.01, 0.1, 0.7$.
Dashed lines are $x_\infty/L$ curves 
as given by Eq.~(\ref{xInfty}). 
All parameters are unity unless otherwise stated.
\label{coupMorph_xtOverL_Fig0}
}
\end{figure}

Let us assume that the boundary between different gene expression regions is determined by the position $x_t$ at which $A$ equals some threshold value $A_t$. 
Inverting, the expression for the threshold position is 
\beq
x_t  (L) = L - \lambda_a \cosh^{-1}\left(A_t/A(L)\right).
\label{xtLowEq}
\eeq
Note that there is a minimum system size for a specific threshold,
\beq
L_m = \lambda_a \sinh^{-1}\left(\frac{\lambda_a \Gamma_a }{D_a A_t}\right) ,
\eeq
such that $x_t(L_m)=L_m$. When $L-x_t \gg \lambda_a$ the concentration profile becomes 
purely exponential and 
$x_t \rightarrow x_\infty$ where
\beq
x_\infty = \lambda \ln\left(\frac{\lambda_a \Gamma _a }{D_a A_t}\right).
\label{xInfty}
\eeq
Clearly,  the function $x_t(L)$ starts out at $L_m$ (which is greater than $x_{\infty}$), is monotonically decreasing, but is bounded below  by $x_{\infty}$; in other words $x_t$ is always greater than  $x_\infty$ as the effect of the zero-flux boundary condition is to make $x_t$
larger than it would be in the absence of the boundary. 
Fig.~\ref{coupMorph_xtOverL_Fig0} shows the variation of $x_t/L$ with $L$ for three different values of the threshold concentration $A_t$. There is no 
extremum where this ratio becomes locally $L$-independent.

\section{Combinatorial model}
\label{combinatorialSect}

We next ask whether a molecular mechanism, which compares the 
concentrations of two gradients rather than reading the absolute value
of one or both of them, can lead to gene expression boundaries which 
scale with system size.
We consider two opposing gradients $A$ and $B$ described by
\beqa
0& = & D_a \partial_x^2 A  - \beta_a A \label{simpledegEq1}\\
0 & = & D_b \partial_x^2 B - \beta_b B \label{simpledegEq2}
\eeqa
at steady state. The boundary conditions are
\beq
\begin{array}{ll}
-D_a\partial_x A (0) = \Gamma_a; & -D_a\partial_x A (L) = 0 \\
-D_b\partial_x B (0) = 0; & -D_b\partial_x B(L) = -\Gamma_b .
\end{array}
\eeq
In this scenario, the gene expression boundary will be determined by a critical concentration ratio $r$ which occurs at the position  $x_r$ defined by $A(x_r) = rB(x_r)$. 

Just as in the one-gradient case, one can distinguish between relatively small systems (for which the no-flux boundary conditions matter) and large systems, depending on how big $L$ is compared to the decay lengths $\lambda_i$. 
For sufficiently large $L$, the gradients of $A$ and $B$ 
are purely exponential, $A = A(0) \exp(-x/\lambda_a)$ and 
$B = B(L) \exp(-(L-x)/\lambda_b)$, and the gene expression boundary 
is given by 
\beq
x_r = \frac{\lambda_a}{\lambda_a+\lambda_b} \left( L \pm L_c \right)
\label{xrExp}
\eeq
where
\beq
L_c=\lambda_b 
\left|\ln\left(\frac{rB(L)}{A(0)}\right)\right| .
\eeq
Consider a gene whose cis-regulatory module contains overlapping $A$ and $B$ binding sites. This gene will have a particular threshold ratio $r$ and a concomitant value $L_c(r)$. Then for sufficiently large developing fields $L\gg L_c(r)$ 
the combinatorial mechanism sets the boundary of expression of the gene at the relative location 
\beq
\frac{x_{r}}{L} =\frac{\lambda_a}{\lambda_a+\lambda_b} 
\eeq
in a size-invariant manner. This position is also insensitive to source-level fluctuations, which only enter in $L_c$.. In a system in which the degradation lengths of the two morphogen gradients are comparable, $x_{r}/L$ will be close to $1/2$. 

\begin{figure}
\centerline{\includegraphics[scale=0.8]{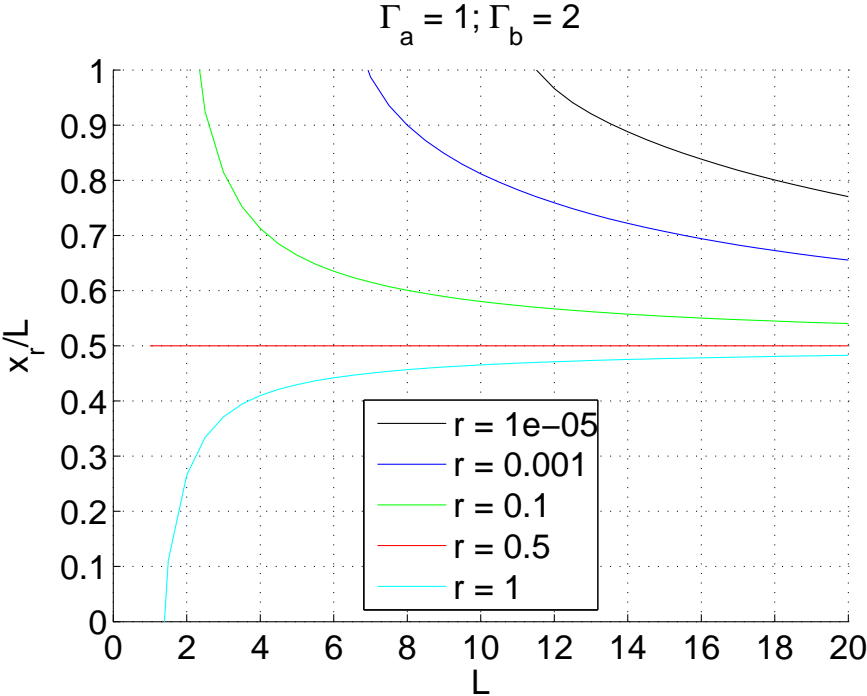}}
\caption{Dependence of normalised $x_r$ on system length $L$
in the combinatorial two-gradient model, as given by Eq.~(\ref{finiteSystemEqCombinatorial}), for values of the threshold ratio equal to (from top to bottom) $r=10^{-5},10^{-3},10^{-1},0.5,1$. All parameters are unity unless otherwise stated.
\label{xrOverLCombinatorial}
}
\end{figure}

Although this model can achieve some degree of size-scaling near the centre of the developing field, from Eq.~(\ref{xrExp}) it is clear that the variation of $x_r/L$ with $L$ increases as $x_r/L$ deviates from the aforementioned asymptotic value. This will happen either as the size of the system is made smaller, or even at fixed $L$ if we try to make the threshold point $x_r$ approach the edges of the developing field. We have in mind a situation where multiple genes need to be regulated, each at different points along the developing field; each gene will have its own value of $r$  and hence its own value of $x_r$. In the previous limit, there is no variation in $x_r$ with $r$ and this cannot be accomplished; therefore we need to rely on finite $L_c/L$ effects. To proceed, we must more carefully characterize the variation of $x_r/L$ with $L$ for all positions in the developing field. Since Eq.~(\ref{xrExp}) becomes inaccurate close to the edges of the developing field we return to the expression for $A$ in Eq.~(\ref{obvEq}) (and a similar one for $B$) and obtain the following implicit equation for $x_r$
\beq
\frac{A(L)}{\cosh(x_r/\lambda_a)} = \frac{rB(0)}{\cosh((L-x_r)/\lambda_b)}
\label{finiteSystemEqCombinatorial}
\eeq
valid  for a finite system. It will be critical to identify what happens to $x_r$ when the length $L$ is made smaller. Notice that there is a different behavior depending on which of $A(L)$ and $rB(0)$ is larger. Specifically, if $A(L)$ is larger, there will be a smallest length below which $x_r$ given by this formula becomes larger than $L$; this length is given by
$$ L^*(r) = \lambda _a \cosh ^{-1} \left( \frac{A(L)}{rB(0)} \right). $$ If, on the other hand, the ordering is reversed, below the length scale
$$ L^*(r) = \lambda _b \cosh ^{-1} \left( \frac{rB(0)}{A(L)} \right) $$ we obtain negative values for $x_r$. Representative $x_r/L$ curves are shown in Fig.~\ref{xrOverLCombinatorial} for the case of equal decay lengths $\lambda_a=\lambda_b$. 

\begin{figure}
\begin{center}
\begin{tabular}{c}
\includegraphics[scale=0.8]{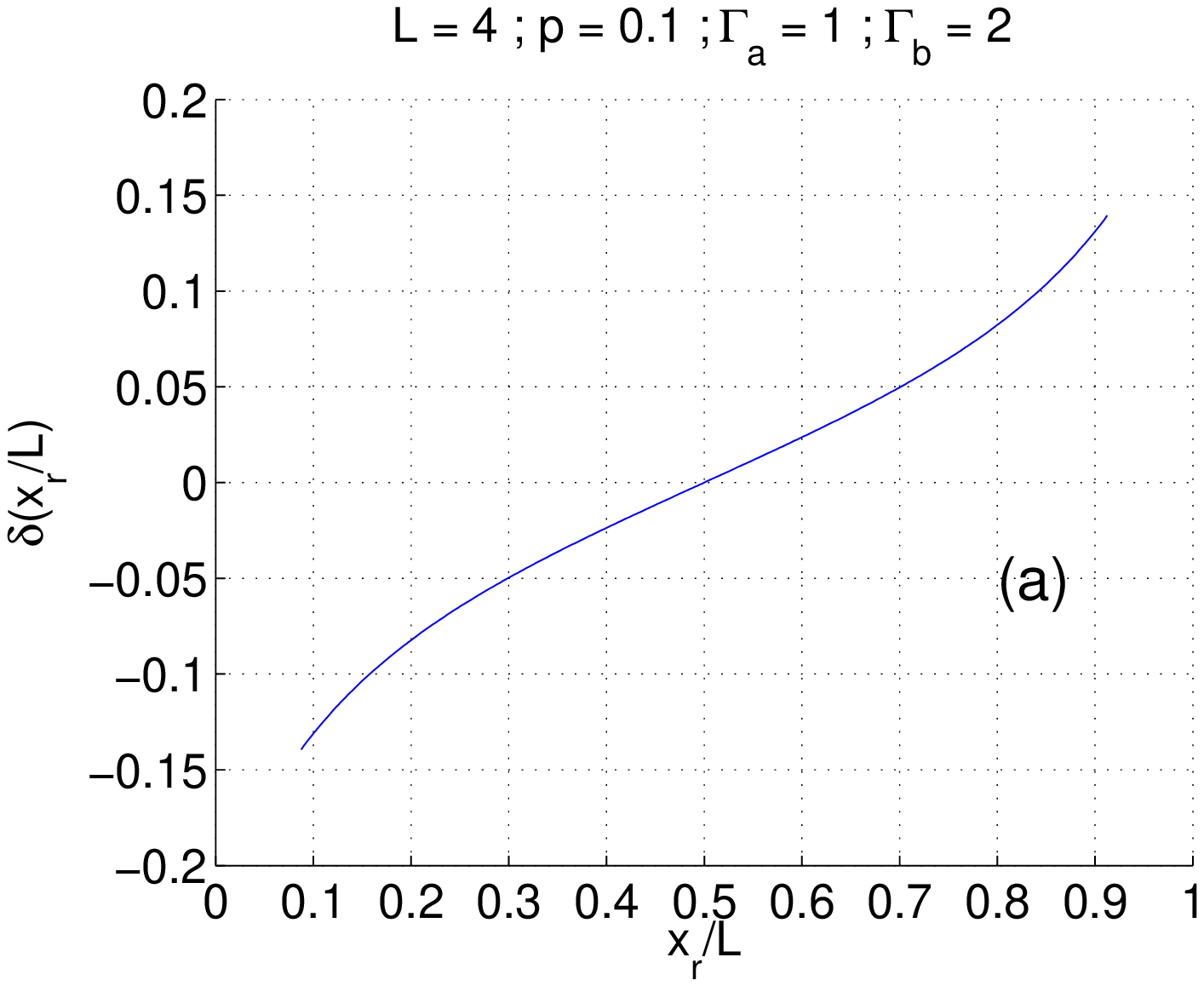}
\\
\includegraphics[scale=0.8]{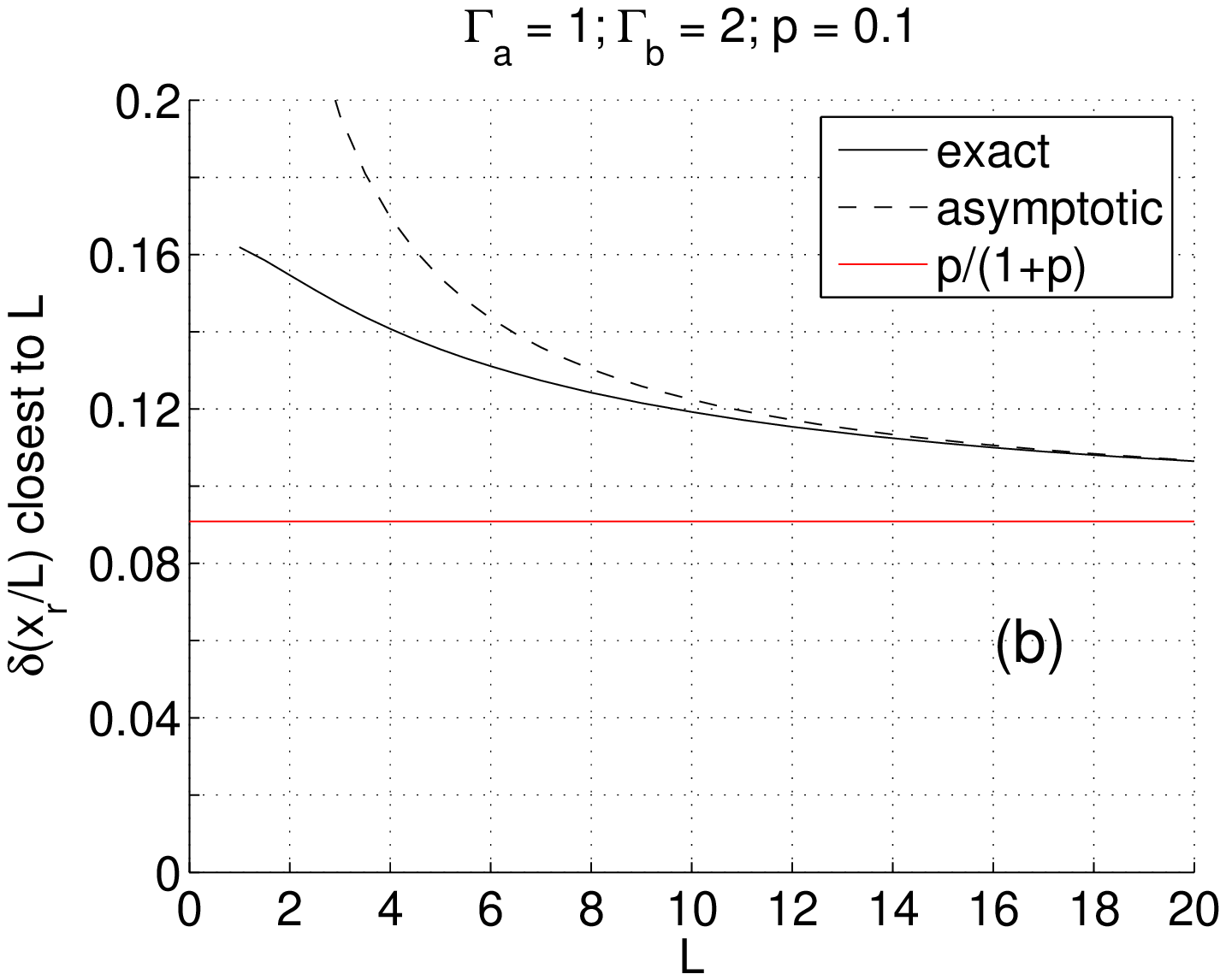}
\end{tabular}
\end{center}
\caption{(a) Dependence of the variation $\delta(x_r/L)$ on normalised position $x_r/L$ in the developing field for $L=4$ and a percentage change in system size of $10\%$.  (b)  The variation $\delta\left(x_{r}/L\right)$ closest to the right boundary of the developing field as a function of $L$ (solid line). At each $L$ we have chosen the target gene whose threshold ratio $r$ satisfies $L^*(r) = L - pL$. The variation in the fractional position at which this gene is turned on is then given by 
$
\delta\left(\frac{x_{r}}{L}\right) = 1 - \frac{x_{r}(L+pL)}{L+pL}.
$ The dashed line is an asymptotic expression. The horizontal (red) line is the limiting value $p/(1+p)$ of the solid and dashed curves.
\label{delta_xrOverL}
}
\end{figure}

Consider now a developing field of size $L$ subject to a natural variation in size of $L\pm pL$ with $0\leq p \leq 1$. The variation in the fractional position at which a gene is turned on is then given by 
\beq
\delta\left(\frac{x_{r}}{L}\right) \equiv  \frac{x_{r}(L-pL)}{L-pL}- \frac{x_{r}(L+pL)}{L+pL}.
\eeq
We show in Fig.~\ref{delta_xrOverL}(a), again for the equal decay length case,  the dependence of $\delta\left(x_{r}/L\right)$ on normalised position $x_r/L$ in the developing field for $L=4$. As expected the variation is largest (in magnitude) at the boundaries and vanishes at that position $x_{r_0}$ for which the critical length $L_c(r_0)$ vanishes. Defining an arbitrary scaling criterion according to 
\beq
\delta\left(x/L\right) \leq 5\%,
\label{scalCrit}
\eeq
one sees that the combinatorial model achieves scaling only in the central region of the developing field between about $30\%$ and $70\%$ of $L$. 

Near the edges of the developing field the variation $\delta\left(x_{r}/L\right)$ is about $14\%$. Since the slopes of the $x_r/L$ curves at $x_r/L=1$ become flatter as $L$ is increased (see Fig.~\ref{xrOverLCombinatorial}), one might wonder whether operating at larger system sizes will decrease this variation. However at larger system sizes the flattening effect is offset by the fact that one must sample larger and larger portions of the $x_r/L$ curve. The extent to which these effects cancel is shown in Fig.~\ref{delta_xrOverL}(b) where we show the variation $\delta\left(x_{r}/L\right)$ closest to the right boundary of the developing field as a function of $L$. The variation decreases with $L$, but an elementary calculation reveals that it has the lower bound $p/(1+p)$. For a percentage variation $p=10\%$ in system size this lower bound is about $9\%$. We conclude that increasing system size is not sufficient to make the combinatorial model, with $\lambda_a=\lambda_b$, meet the scaling criterion throughout the developing field.

A further difficulty with the  combinatorial model is its susceptibility to small-molecule-number fluctuations. In general, we must expect $L_c$ of order $\lambda$, since we cannot independently adjust the morphogen sources for the multiple genes that need to be controlled. In fact, the natural interpretation of $r$ as being due to binding differences between different transcription factors suggest that $L_c$ would vary significantly.  In such cases the limit $L\gg L_c(r_0)$ would force the comparison point $x_r$ far down the profile from the source; having enough molecules at this point to affect the necessary DNA binding would then place a severe constraint on source strengths. 
In this regard a combinatorial mode of action may
favour power-law (resulting e.g. from nonlinear degradation~\cite{Eldar2003})
 over exponential profiles as the 
former have greater range than the latter, but this remains to be studied.

\section{Annihilation model}
\label{coupledSect}

We return to the standard model of morphogenesis in which cell-fate
boundaries are determined according to the position at 
which a single morphogen crosses a threshold concentration.
We couple this gradient to an auxilary gradient directed from 
the opposite end of the developing field. We then ask under what conditions
the primary gradient may scale with system size. 

We  consider two species of morphogen, $A$ and $B$, 
in a one-dimensional system of length L
with $A$s and $B$s injected at opposite ends of the system. 
The boundary
conditions are as in \S~\ref{combinatorialSect}.
 The 
species interact according to the annihilation reaction 
$A+B\rightarrow\emptyset$. In a mean-field description the 
kinetics is described by the reaction-diffusion equations
\beqa
\partial_t A & = & D_a \partial_x^2 A -\beta_a A
- kAB \label{Redner1}\\
\partial_t B & = & D_b \partial_x^2 B -\beta_b B
- kAB \label{Redner2}
\eeqa
where $k$ is 
the annihilation rate constant. Later, we will consider more complex models which incorporate non-linear degradation or non-linear (i.e. concentration-dependent) diffusion.

This system of equations, with fluxes $\Gamma_a=\Gamma_b=\Gamma$ and without any decay,
was considered by Ben-Naim and Redner \cite{BenNaim1992}. They 
determined the steady-state spatial distribution of the reactants and 
of the annihilation zone which they 
chose to be centred in the interval $[0,L]$. The annihilation zone is roughly 
the support of 
$R(x)=kA(x)B(x)$ or, put another way, that region 
where the concentration of both species is appreciable. With the aid 
of a rate-balance argument, they showed that 
the width $w$ of the 
annihilation zone scales as $\Gamma^{-1/3}$ and
that the concentration in this zone is proportional to $\Gamma^{2/3}$
when   $w \ll L$.

Our goal is to understand the relation of the steady-state 
concentration profiles to the 
system length $L$. It is convenient to 
identify the point  $x_e$ in the annihilation zone where 
the profiles cross, $A(x_e)=B(x_e)$. In the original Ben-Naim---Redner model, 
the reaction-diffusion
equations yield no unique value for $x_e$;
instead $x_e$ can lie anywhere in the interval $[0,L]$ depending 
on the choice of initial condition.
To see this consider the following rate-balance argument.
Since the particles annihilate in 
a one-to-one fashion the flux of each species into the annihilation 
zone must 
be equal. But this condition does not determine $x_e$ uniquely because
these fluxes are always 
equal to the input fluxes at the boundaries. Similarly, the model without degradation cannot support steady states with unequal boundary fluxes.
If, however,
we now add degradation terms to the steady-state equations,
then the flux of each species into the annihilation zone is the 
flux into the system less the number of degradation events that happen before reaching the zone. Thus,
the flux of each species into the annihilation zone now depends on the location 
$x_e$ and so there is only one value of $x_e$ which balances the fluxes. As we will see, our models will always contain unique steady-state solutions.

A rough estimate of the concentration in the annihilation zone and 
of the width of the zone can be obtained using the 
original Ben-Naim--Redner rate-balance argument \cite{BenNaim1992}.
We identify three spatial regions: the first where $A$ is in the majority;
the second the annihilation zone; and the third where 
$A$ is in the minority. Assume the concentration of $A$s in this latter region
is negligible compared with that in the other two regions.
The concentration of $A$s 
in the annihilation zone should then be of the order of the slope of the 
concentration profile 
in the annihilation zone times the width $w$. The slope of 
the $A$ profile in this region is proportional
to $j_e/D_a$, where $j_e$ is the equal flux of $A$s or $B$s 
into the annihilation zone. Therefore the concentration in
the annihilation zone $A_e=A(x_e)$ is 
\beq
A_e \sim j_e w/D_a.
\label{AeEQ}
\eeq
If we ignore the loss of $A$ particles in the 
annihilation zone (valid for small $w$), then the number of annihilation 
events per unit time $kA_e^2w$ should equal  
the flux $j_e$. Balancing these two rates gives
$j_e \sim k(j_e w /D_a)^2 w $. Hence the width of 
the annihilation zone scales as
\beq
w \sim \left(\frac{D_a^2}{j_e k}\right)^{1/3}.
\label{wScalingFunction1}
\eeq
In what follows, we will be mostly interested in taking $k$ large enough to give a very small $w$.

\section{The high-annihilation-rate limit}
\label{sectHighAnnLimit}

We now explicitly assume that the parameters lie in the limit where $w \ll \mbox{min}\{x_e,L-x_e\}$.
This limit has the considerable advantage that 
the $A$-$B$ system may be decoupled by replacing the coupling 
term $kAB$ by a zero-concentration
boundary condition at $x_e$. 
In this approximation the concentration 
of the $A$ subsystem 
satisfies 
\beq
0  =  D_a \partial_x^2 A - \beta_a A 
\eeq
subject to the boundary conditions
$
-D_a\partial_x A (0) = \Gamma_a$ and $  A (x_e) = 0$ .
The solution to this equation is
\beq
A(x) = \frac{\lambda_a \Gamma _a }{D_a \cosh(x_e/\lambda_a)} \sinh\left(\frac{x_e-x}{\lambda_a}\right) = A_*\sinh\left(\frac{x_e-x}{\lambda_a}\right) 
\label{AEqLinDeg}
\eeq
 where as before
$\lambda_a
= \sqrt{D_a/\beta_a}$. $A_*$ is a characteristic concentration of the 
$A$ field related to the slope of the $A$ field at $x_e$ according to
$A_* = -\lambda_a \partial_x A(x_e) .$
The flux of $A$ particles is 
\beq
j_a(x)=j_a(x_e) \cosh\left(\frac{x_e-x}{\lambda_a}\right) .
\label{fluxAnnZoneEq}
\eeq
where the  flux into the 
annihilation zone $j_a(x_e)$ is given by $j_a(x_e) =\Gamma_a / 
\cosh\left(x_e/\lambda_a\right)$. Substituting this 
into Eq.~(\ref{wScalingFunction1}) yields
the scaling function of the annihilation 
zone width for the case of linear degradation
\beq
w \sim w_0 \left[ \cosh(x_e/\lambda_a) \right]^{1/3}.
\label{wEqLin}
\eeq
Here 
$
w_0 \sim \left(D_a^2/\Gamma_a k\right)^{1/3}
$
is the width of the annihilation zone  in the absence of degradation 
\cite{BenNaim1992}.
Note that we may also substitute this expression for $j_a(x_e)$ into
Eq.~(\ref{AeEQ}) obtaining $A_e\sim w /  \cosh(x_e/\lambda_a)$. One can then verify that $A_e$ is much smaller than $A(0)$ whenever $w \ll x_e$ and hence approximating this as a zero boundary condition is self-consistently valid.

The $B$-subsystem can be treated similarly, except that the length of the
subsystem in this case is $L-x_e$. The only dependence on the annihilation 
rate $k$ in the inequality $w \ll x_e$ 
occurs in $w_0$.  Hence this limit is equivalent to the
 high-annihilation-rate limit 
$k \gg k_0$, where the threshold value $k_0$ of the annihilation rate is given by
\beq
k_0 \sim \frac{D_a^2}{\Gamma_a \lambda_a^3} \frac{\cosh(x_e/\lambda_a)}
{(x_e/\lambda_a)^3}. 
\eeq

\begin{figure}
\begin{center}
\begin{tabular}{c}
\includegraphics[scale=0.8]{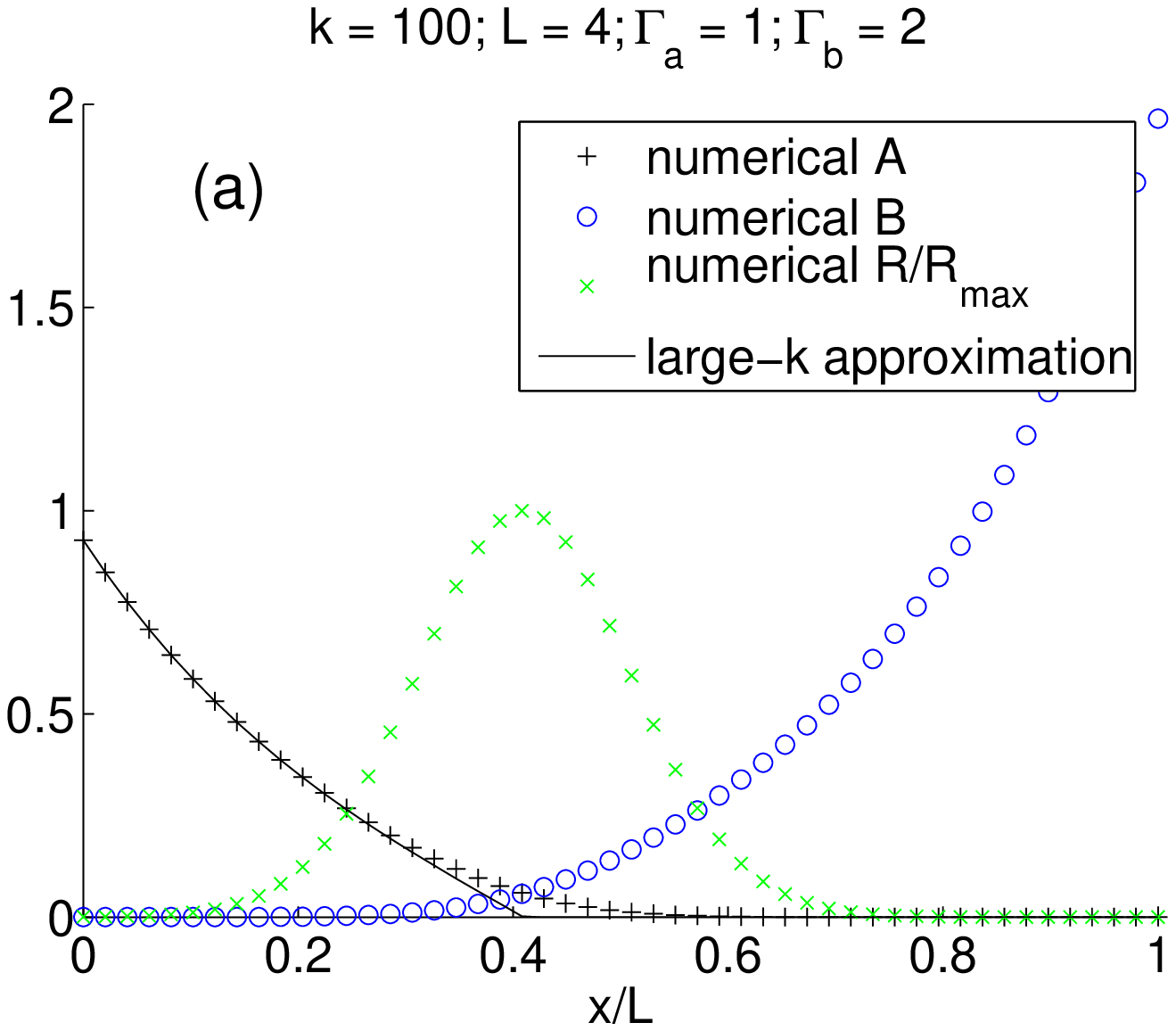}
\\
\includegraphics[scale=0.8]{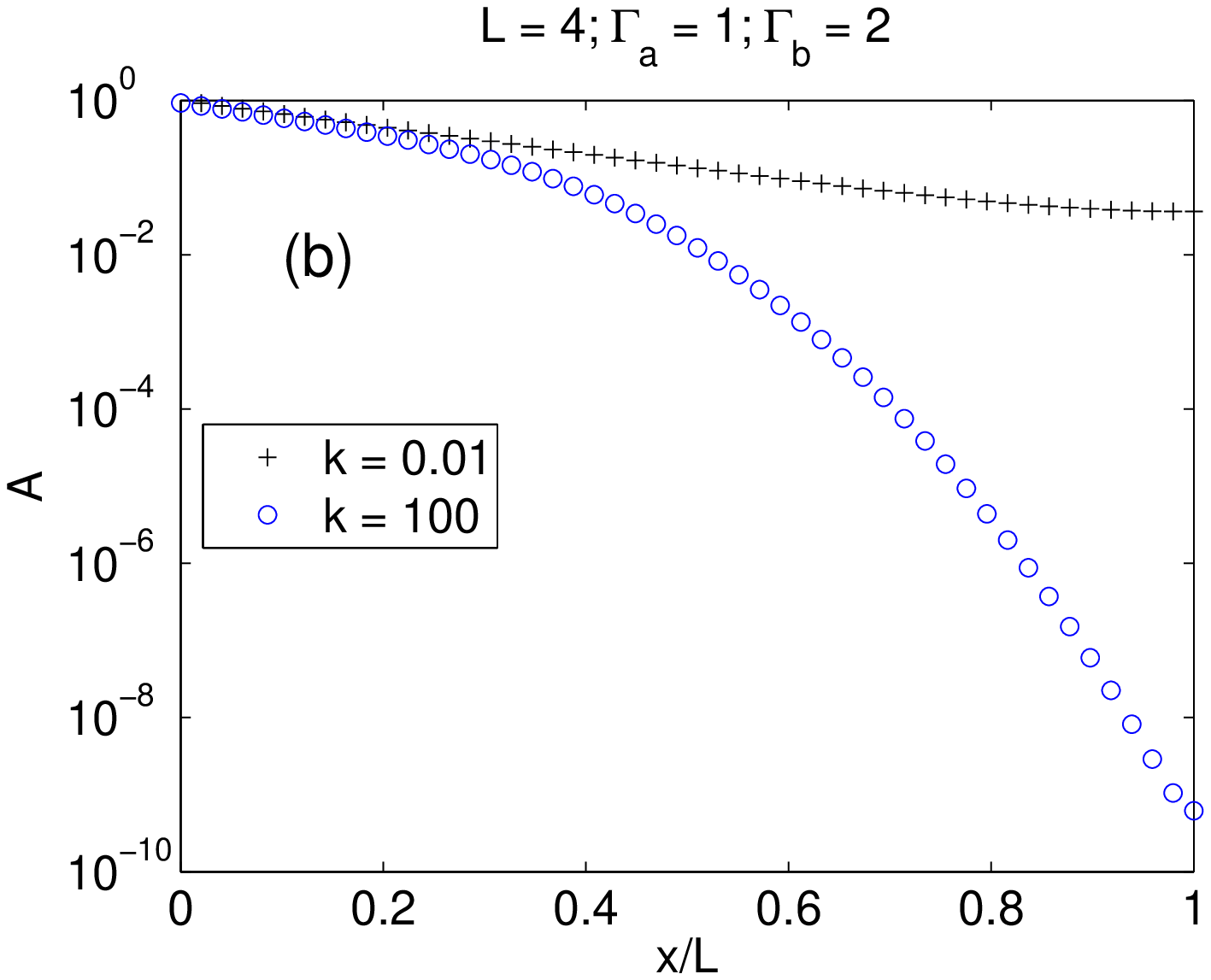}
\end{tabular}
\end{center}
\caption{(a) Comparison of the high-annihilation-rate 
approximation (Eq.~(\ref{AEqLinDeg}), solid line) with the numerical solution (circles)
of the full annihilation model (Eqs.~(\ref{Redner1}) and~(\ref{Redner2})). The annihilation zone is the reaction front $R(x) = kA(x)B(x)$. 
All parameters are unity unless otherwise stated.
(b) A(x) plotted on a logarithmic 
scale in the cases $k=0.01$ and $k=100$. Note the crossover from slow decay
 in the $A$-rich region to fast decay in the 
$B$-rich region in the case $k=100$.
All parameters are unity unless otherwise stated.
\label{coupMorph_xtOverL_Fig5}
}
\end{figure}

We determine the annihilation zone location by balancing fluxes into the zone, 
$j_a(x_e) = - j_b(L-x_e)$. This leads to the following equation for $x_e$
\beq
\frac{\Gamma_a}{\cosh\left(\frac{x_e}{\lambda_a}\right)} = \frac{\Gamma_b}{\cosh\left(\frac{L-x_e}{\lambda_b}\right)}.
\label{eqFluxes}
\eeq
In the special case $\lambda_a=\lambda_b$ this equation coincides with the implicit definition of $x_r$ (with $r=1$)  which arose in the combinatorial model (see Eq.~(\ref{finiteSystemEqCombinatorial})). As in that model there is a smallest length $L^*$ defined by $$ L^* = \lambda _a \cosh ^{-1} \left( \frac{\Gamma_a}{\Gamma_b} \right)$$ if $\Gamma_a>\Gamma_b$ and by $$ L^* = \lambda _b \cosh ^{-1} \left( \frac{\Gamma_b}{\Gamma_a} \right) $$ if the flux ordering is reversed. As our entire treatment of the annihilation zone only makes sense if $ 0 \leq x_e \leq L$, we must always choose $L >L^*$. A comparison of the numerical solution of the full model with the results of the large-annihilation-rate approximation is shown in Fig.~\ref{coupMorph_xtOverL_Fig5}.

Once we know $x_e(L)$ and $A(x)$,
we can proceed to determine 
the qualitative features of the $x_t(L)$ function with a view
to identifying the region of system sizes where $x_t \sim L$.
Inverting Eq.~(\ref{AEqLinDeg}) we find 
\beq
x_t = x_e - \lambda_a \sinh^{-1}\eta
\label{xtEqLinDeg}
\eeq
where 
\beq
\eta  = A_t / A_* .
\label{etaEq}
\eeq
Note that $x_t$ depends on $L$ only through its dependence on $x_e$ and the function $x_t (x_e)$ is monotonically increasing.
Obviously $x_t \leq  x_e$. In the limit of sufficiently large $x_e$, we can replace the inverse hyperbolic function with a logarithm and obtain the simpler form
\beq
x_t \approx x_e - \lambda_a \ln(2\eta).
\label{xtApprox}
\eeq
Here, $\eta \approx \frac{A_t}{A(0)}\frac{1}{2} e^{x_e/\lambda_a}$, and $x_t$ approaches its asymptotic value $x_\infty \approx \lambda_a \ln(A(0)/A_t)$ from below. This is of course the answer one would obtain in the absence of any auxiliary gradient.

\begin{figure}
\centerline{\includegraphics[scale=0.8]{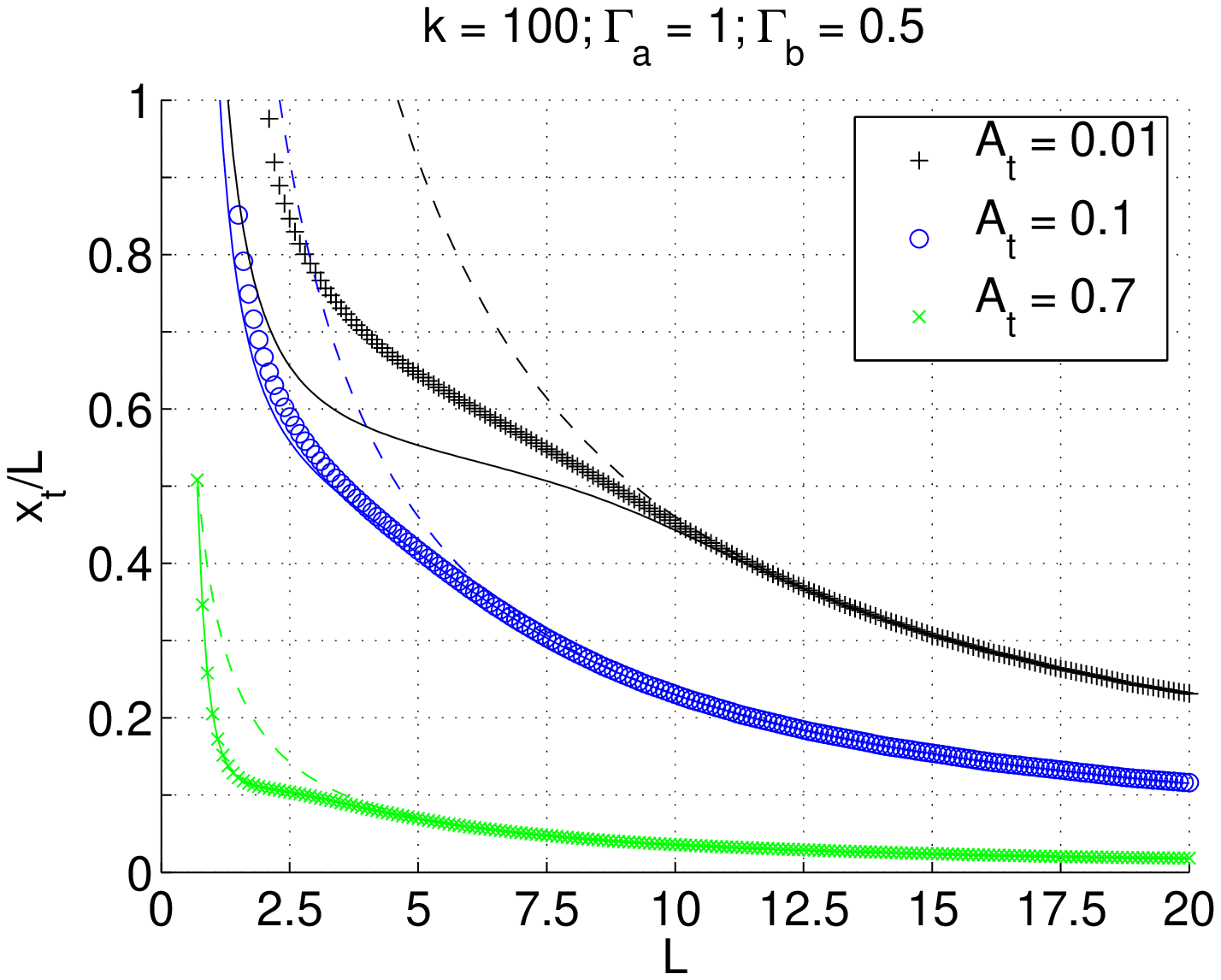}}
\caption{Dependence of normalised $x_t$ on system length $L$
with $k=100$ and $\Gamma_b < \Gamma_a$. 
The plus signs, circles and crosses 
are numerical solutions of Eqs.~(\ref{Redner1}) 
and~(\ref{Redner2}) for values of the 
threshold concentration equal to (from top to bottom) $A_t=0.01, 0.1, 0.7$.
The solid lines  are the 
corresponding analytic expressions Eq.~(\ref{xtEqLinDeg})
obtained in the 
high-annihilation-rate limit.
Dashed lines are $x_\infty/L$ curves 
as given by Eq.~(\ref{xInfty}).  
All parameters are unity unless otherwise stated.
\label{coupMorph_xtOverL_Fig2a}
}
\end{figure}

\begin{figure}
\centerline{\includegraphics[scale=0.8]{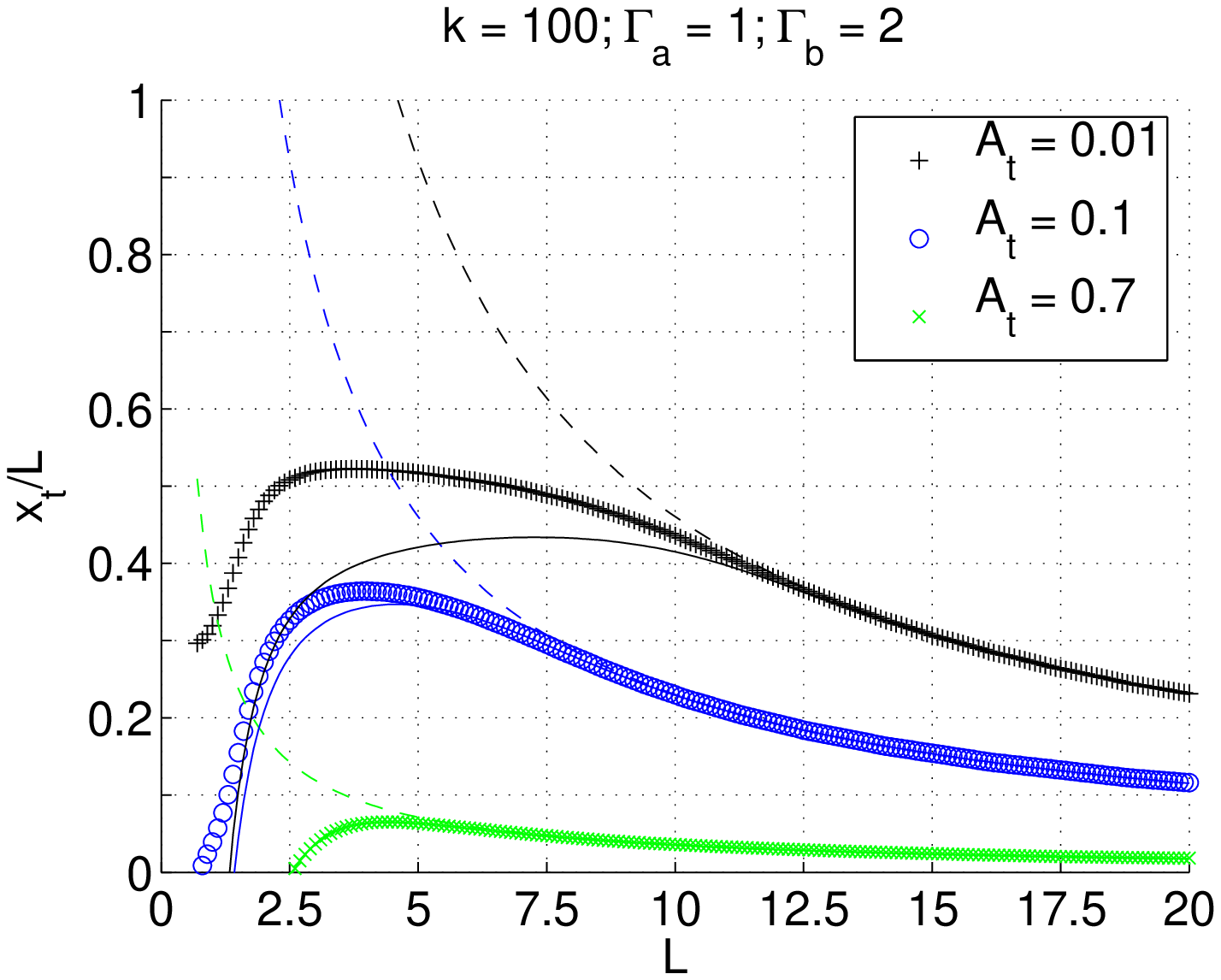}}
\caption{Dependence of normalised $x_t$ on system length $L$
with $k=100$ and  $\Gamma_b > \Gamma_a$. 
The plus signs, circles and crosses are numerical solutions of Eqs.~(\ref{Redner1}) 
and~(\ref{Redner2}) for values of the 
threshold concentration equal to (from top to bottom) $A_t=0.01, 0.1, 0.7$.
The solid lines  are the 
corresponding analytic expressions Eq.~(\ref{xtEqLinDeg})
obtained in the 
high-annihilation-rate limit.
Dashed lines are $x_\infty/L$ curves 
as given by Eq.~(\ref{xInfty}). 
All parameters are unity unless otherwise stated.
\label{coupMorph_xtOverL_Fig4}
}
\end{figure}

\begin{figure}
\centerline{\includegraphics[scale=0.8]{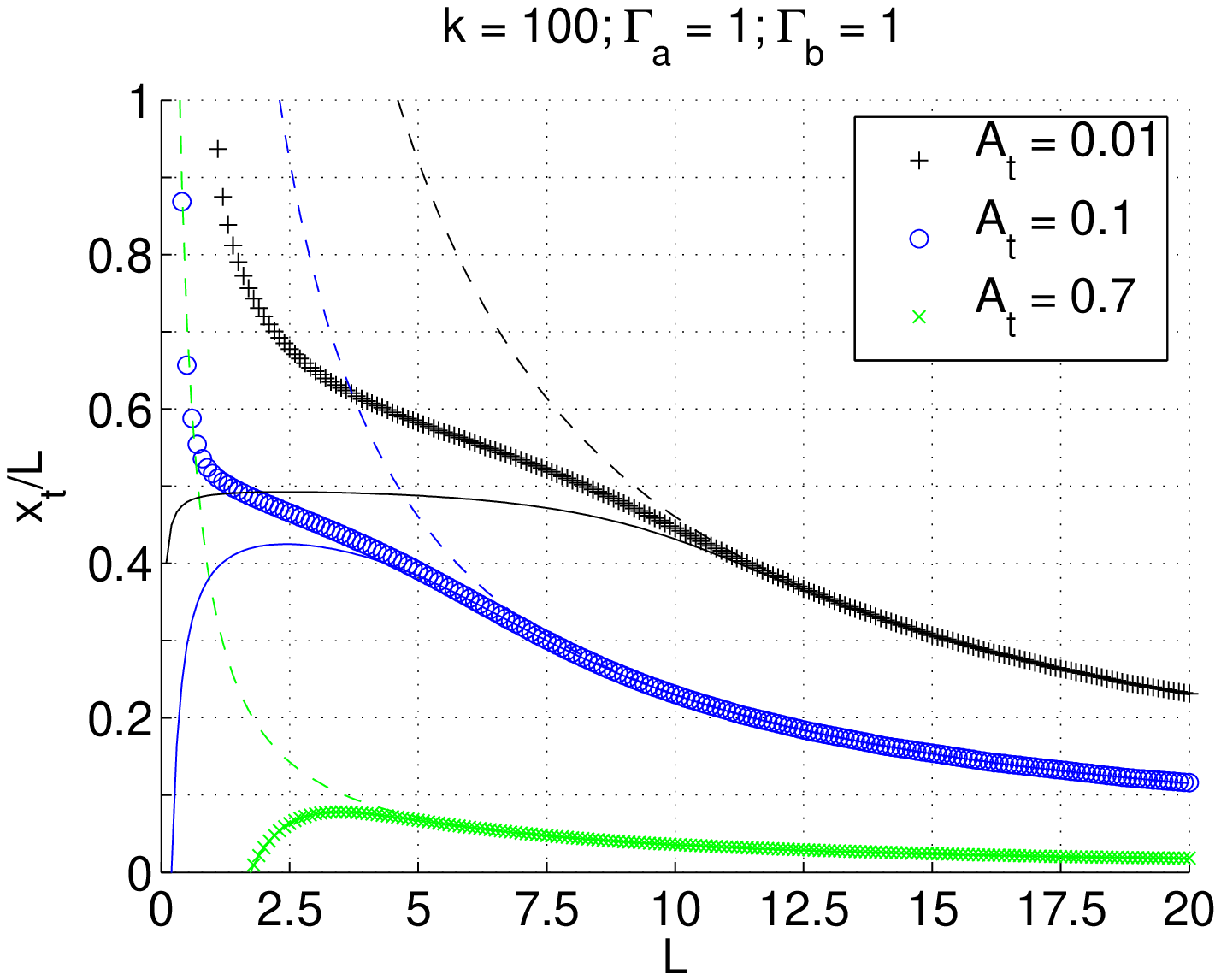}}
\caption{Dependence of normalised $x_t$ on system length $L$
with $k=100$ and $\Gamma_b = \Gamma_a$. 
The plus signs, circles and crosses are numerical solutions of Eqs.~(\ref{Redner1}) 
and~(\ref{Redner2}) for values of the 
threshold concentration equal to (from top to bottom) $A_t=0.01, 0.1, 0.7$. 
The solid lines  are the 
corresponding analytic expressions Eq.~(\ref{xtEqLinDeg}) obtained in the 
high-annihilation-rate limit.
Dashed lines are $x_\infty/L$ curves 
as given by Eq.~(\ref{xInfty}). 
All parameters are unity unless otherwise stated.
\label{coupMorph_xtOverL_Fig3}
}
\end{figure}

Now, imagine reducing $L$ and hence $x_e$ from its just-mentioned asymptotic regime and plotting the ratio $x_t/L$. For the case $\Gamma_a > \Gamma_b$, $x_e$ will eventually hit $L$ followed shortly thereafter by $x_t/L$ hitting unity. There is no reason why this curve should exhibit a maximum, and a direct numerical calculation for $k=100$ (shown in Fig.~\ref{coupMorph_xtOverL_Fig2a}) verifies this assertion. The situation is dramatically different, however, for the case of  $\Gamma_b> \Gamma_a$. Now $x_e$ must approach zero, implying that at some larger $L$ we have $x_t$=0. The curve $x_t/L$ now exhibits a maximum, as is again verified by direct numerical calculations using both the large-annihilation-rate approximation and also just solving the initial model with no approximations whatsoever (see Fig.~\ref{coupMorph_xtOverL_Fig4}). Near the peak of the curve we have scaling with system size. For completeness, we also present in Fig.~\ref{coupMorph_xtOverL_Fig3} the results for equal fluxes.
 
\begin{figure}
\centerline{\includegraphics[scale=0.8]{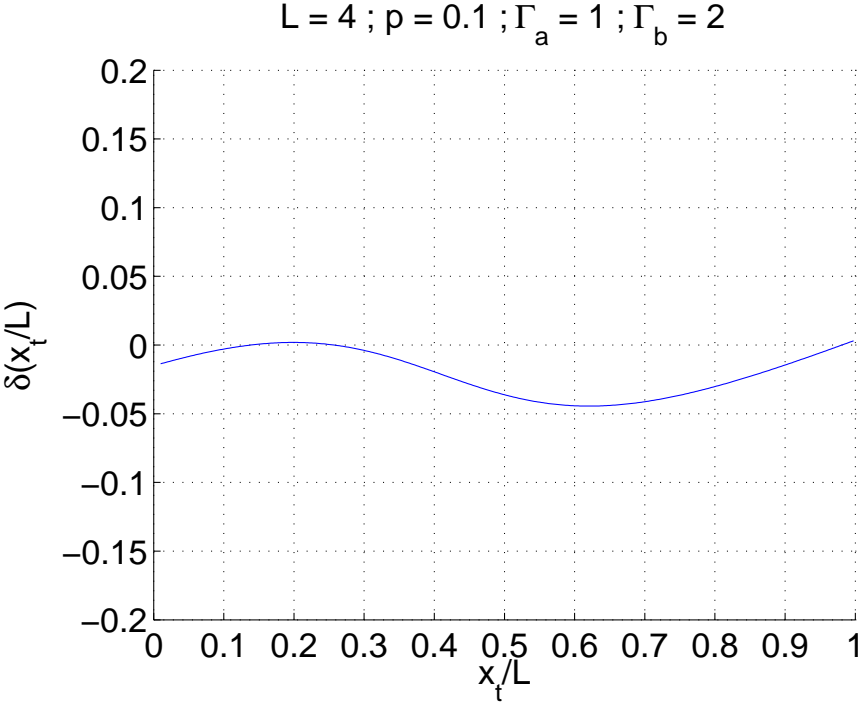}}
\caption{The dependence of the variation $\delta\left(x_t/L\right) $ on normalised position $x_t/L$ in the high-annihilation-rate approximation of the annihilation model. Positions to the left of $x_e$ are set by the $A$ gradient while positions to the right are set by the $B$ gradient. 
All parameters are unity unless otherwise stated.
\label{delta_xtOverL}
}
\end{figure}

To compare the scaling performance of the annihilation model with that of the combinatorial model we show in Fig.~\ref{delta_xtOverL} the dependence of the variation $\delta\left(x_t/L\right) $ on normalised position $x_t/L$ in the developing field for $L=4$. One sees that, according to our scaling criterion in Eq.~(\ref{scalCrit}), the annihilation mechansim can easily set markers scale-invariantly throughout a developing field whose size is a few decay lengths. Furthermore at such system sizes a range of threshold values spanning  two orders of magnitude ($A_t=0.01-0.7$) is sufficient to cover the entire developing field (see Fig.~\ref{coupMorph_xtOverL_Fig3}). Such a modest variation in concentration  makes  the annihilation model less susceptible to small-molecule-number fluctuations than the combinatorial model.

\section{Discussion}

We have considered two scenarios in which a pair of 
oppositely directed morphogen gradients are used to 
set embryonic markers in a size-invariant manner. 
In the simplest scenario, in which  
the gradients interact only indirectly through overlapping DNA-binding sites,
exponentially distributed fields 
achieve perfect size scaling at a normalised position $\lambda_a/(\lambda_a+\lambda_b)$ determined only by the morphogen decay lengths $\lambda_a$ and $\lambda_b$. For equal decay lengths, the accuracy with which this model can set markers size-invariantly decreases as the boundaries of the developing field are approached. At the boundaries the accuracy can be no better than $\delta(x_r/L) = p/(1+p)$ where $p$ is the percentage variation of the field size. 
 In the second model $A$ and $B$ are coupled via the reaction $A+B\rightarrow\emptyset$ and the embryonic markers are set by a single gradient
with the second gradient serving only to provide size
information to the first. In this scenario, it is easy to arrange parameters such that
scaling occurs with an accuracy $\delta(x_t/L)$ better than $5\%$ over the entire developing field for field sizes of only a few decay lengths. 

In practice a given morphogen may play both roles in patterning, setting markers in a strictly concentration-dependent manner at some locations in the developing field and in a combinatorial fashion at other locations \cite{Small2005}. The annihilation model naturally sets markers via the gradient whose source is closest to the marker \cite{Rajewsky2004}, whereas the combinatorial model is better suited to setting markers in the vicinity of the midpoint of the developing field where the variation $\delta(x_r/L)$ is smallest. As the variation $\delta(x/L)$ has a qualitatively different dependence on $x/L$ in either case, a measurement of this curve in a developmental system may distinguish between the mechanisms. 

The origin of the scaling form $f(x/L)$ 
which arises in the strong-coupling  limit of the annihilation model is the 
effective boundary condition $A(x_e)=0$. 
In the case $\Gamma_b > \Gamma_a$ (see Fig.~\ref{coupMorph_xtOverL_Fig4}) 
the $x_t/L$ curve has a 
maximum because at small $L$ ($L\sim L_*$) 
 it tends to zero along with $x_e/L$ 
while at large $L$ ($L\gg L_*$) 
it is bounded above by $x_\infty/L$.
In the $k\ll k_0$ limit, on the other hand,
the zero-concentration effective boundary condition is 
replaced by a zero-flux boundary condition $j_a(L)=0$
which can never induce the $x_t\sim L$ scaling.

This approach makes it clear why the scaling occurs at intermediate values of $L$. Once we reach the non-overlapping limit where the two fields do not effectively communicate, the threshold is set by the $A$ profile alone; we have already seen that this cannot give any scaling. For $L$ too small, the annihilation-zone width $w$ becomes comparable to $x_e$, there is no effective boundary condition and again scaling fails. In fact, if one looks at the expression for $w/x_e$, namely
\beq
\frac{w}{x_e} \sim \frac{w_0}{\lambda_a} \left( \frac{\cosh(x_e/\lambda_a)}{(x_e/\lambda_a)^3}\right)^{1/3} ,
\label{wxeEq}
\eeq
(where $w_0 \sim (D_a^2/\Gamma_a k)^{1/3}$) one sees that the maximum in
$x_t/L$ occurs close to the minimum of $w/x_e$ which is reached at $x_e/\lambda_a
\approx 3$. 

\begin{figure}
\begin{center}
\begin{tabular}{c}
\includegraphics[scale=0.8]{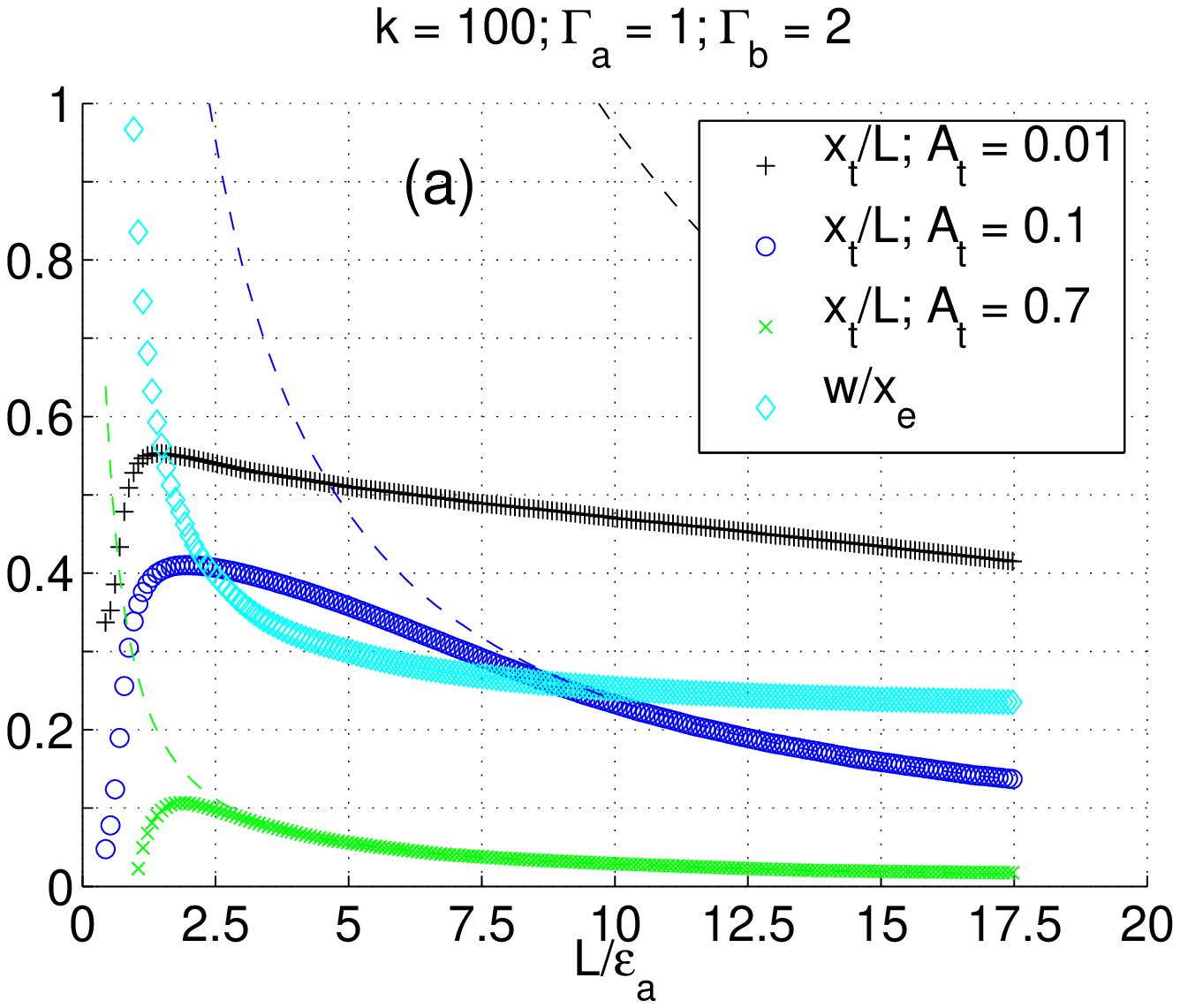}
\\
\includegraphics[scale=0.8]{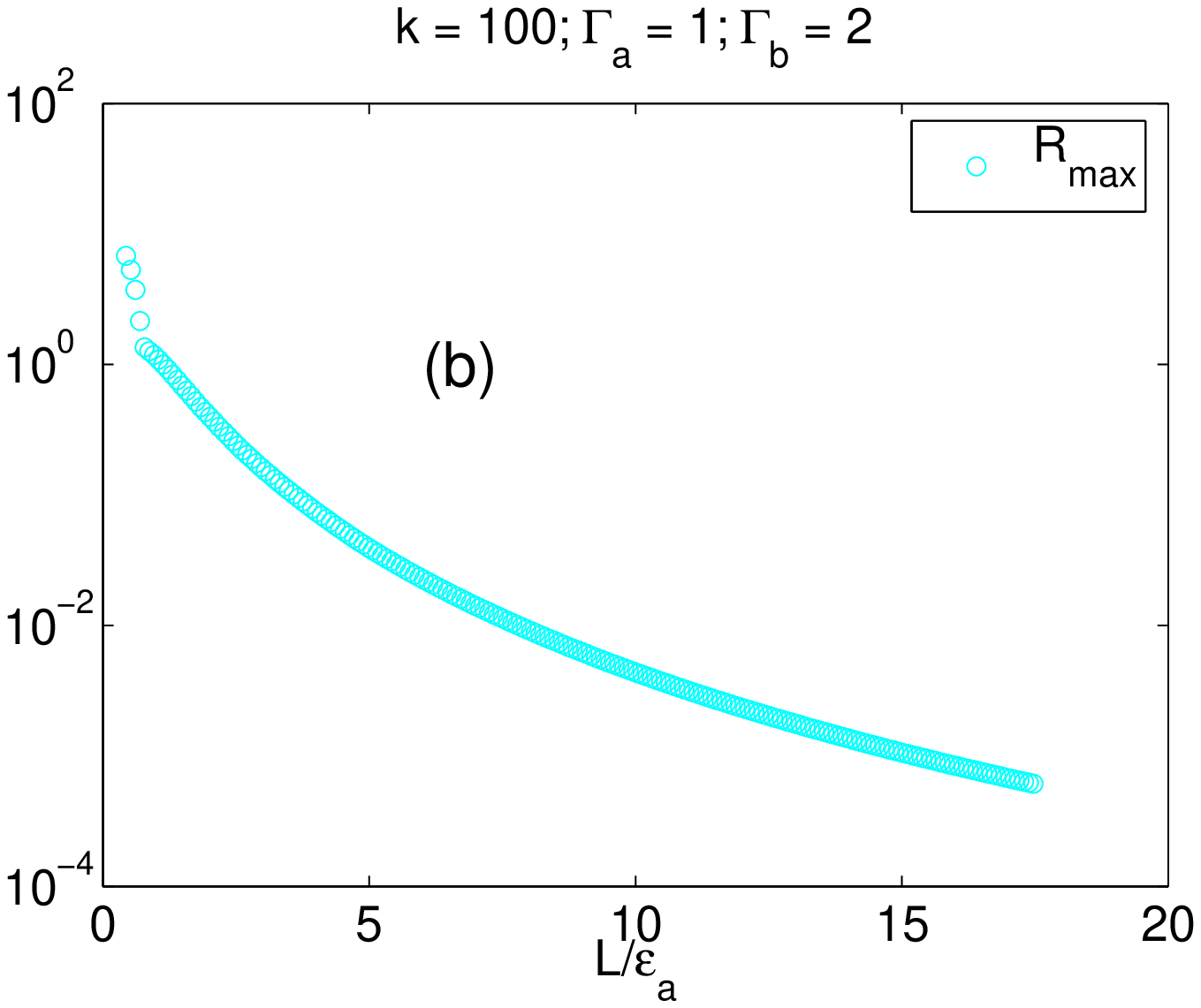}
\end{tabular}
\end{center}
\caption{(a) Dependence of normalised $x_t$ on system length $L$
in the case of quadratic degradation. 
The plus signs, circles and crosses  are numerical solutions 
of the full annihilation  model 
for values of the threshold 
concentration equal to (from top to bottom) $A_t=0.01, 0.1, 0.7$. 
The dashed lines are $x_\infty/L$ curves as 
given by Eq.~(\ref{xInftyQuad}). Also shown (cyan diamonds) 
is the ratio of the 
full width at half maximum $w$ to the comparison point $x_e$. 
(b) The dependence of the amplitude $R_{max}$ of the 
local annihilation rate $R(x)=kA(x)B(x)$ on system length $L$. 
In (a) and (b) all parameters are unity unless otherwise stated. 
\label{coupMorph_xtOverL_Fig9}
}
\end{figure}

\begin{figure}
\centerline{\includegraphics[scale=0.8]{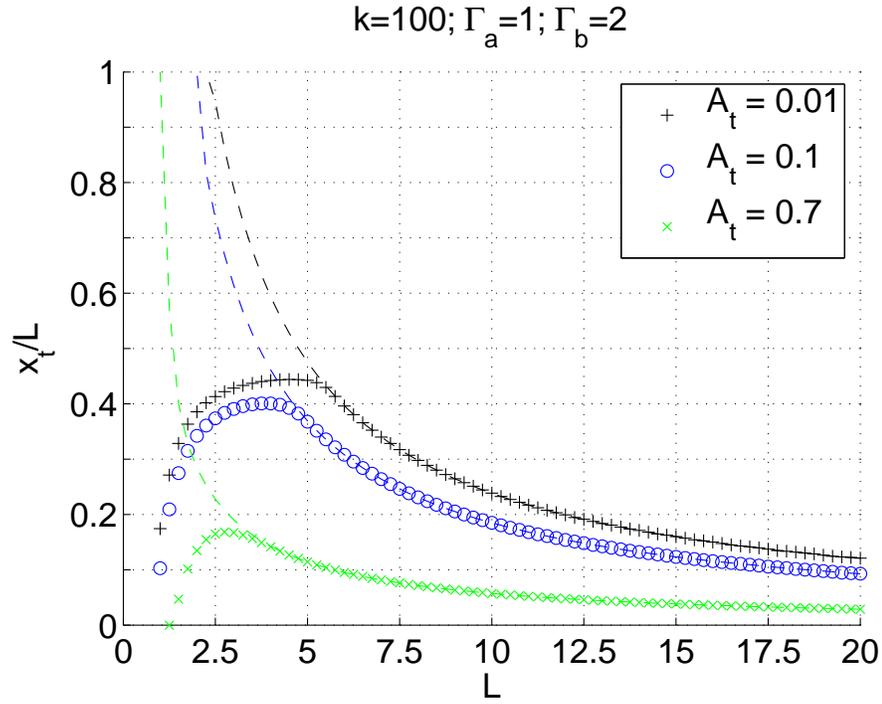}}
\caption{
Dependence of scaled threshold position $x_t/L$ on 
system length $L$ in the simplest case of nonlinear diffusion, 
$D_a = \delta_a A$ and $D_b = \delta_b B$. The degradation terms
are linear. The plus signs, circles and crosses are numerical solutions 
of the full annihilation model 
for values of the threshold
concentration equal to (from top to bottom) $A_t=0.01, 0.1, 0.7$. 
Dashed lines are corresponding curves in the 
case $k=0.01$. All parameters are unity unless otherwise stated. 
\label{nonLinDiffL}
}
\end{figure}

So far we have used linear degradation and simple diffusion in the annihilation model. However, it should be clear from the above arguments that in fact this mechanism is rather robust to changing the nature of the individual gradients. For example, let us consider quadratic degradation. In the limit that the system size is so big as to render the coupling term  $kAB$ irrelevant the $A$ and $B$ profiles reduce to  power laws, $A = a/(x+\epsilon_a)^2$ and  $B = b/(L-x+\epsilon_b)^2$.
The corresponding $L$-independent threshold position $x_\infty$ is given 
by 
\beq
x_\infty = \epsilon_a  \left( \sqrt{\frac{A(0)}{A_t}} - 1 \right) .
\label{xInftyQuad}
\eeq
An argument, similar to one presented earlier for linear degradation, reveals the fact that $x_e$ will be forced to zero for sufficiently small $L$ if $\Gamma_b > \Gamma_a$; this indicates again that to the extent we can believe the large-annihilation-rate approximation, there will be a maximum in the $x_t/L$ curve. This is illustrated for one specific choice of parameters in Fig.~\ref{coupMorph_xtOverL_Fig9}(a). 
The maximum again takes place roughly where $L$ becomes so small as to cause the annihilation-zone width to approach $x_e$. Repeating the derivation of $w$
outlined in \S~\ref{sectHighAnnLimit} but using a power law 
instead of hyperbolic sine we obtain 
\beq
\frac{w}{x_e} \sim \frac{w_0}{\epsilon_a} \left( 1 + \frac{1}{x_e/\epsilon_a} \right).
\eeq
This expression is a good qualitative description of the 
exact $w/x_e$ shown in Fig.~\ref{coupMorph_xtOverL_Fig9}(a) and
diverges when $L \rightarrow 0$ as in the case of linear degradation.
Notice that scaling is lost when $w\rightarrow x_e$ even though 
the rate of the annihilation reaction becomes large (Fig.~\ref{coupMorph_xtOverL_Fig9}(b)). 
Finally, one can also ask about the effect of making the diffusion constant concentration dependent. This type of effect can arise whenever the morphogen reversibly binds to buffers that differ in mobility from the  pure molecule. Fig.~\ref{nonLinDiffL} illustrates the behavior under the simplest assumption, namely that the diffusion constant varies linearly with concentration for both the $A$ and $B$ fields. Aside from sharpening the transition from the asymptotic non-interacting regime to the regime where $x_e$ approaches zero (as $L$ is lowered), the basic phenomenology is unchanged.

\begin{figure}
\begin{center}
\begin{tabular}{c}
\includegraphics[scale=0.8]{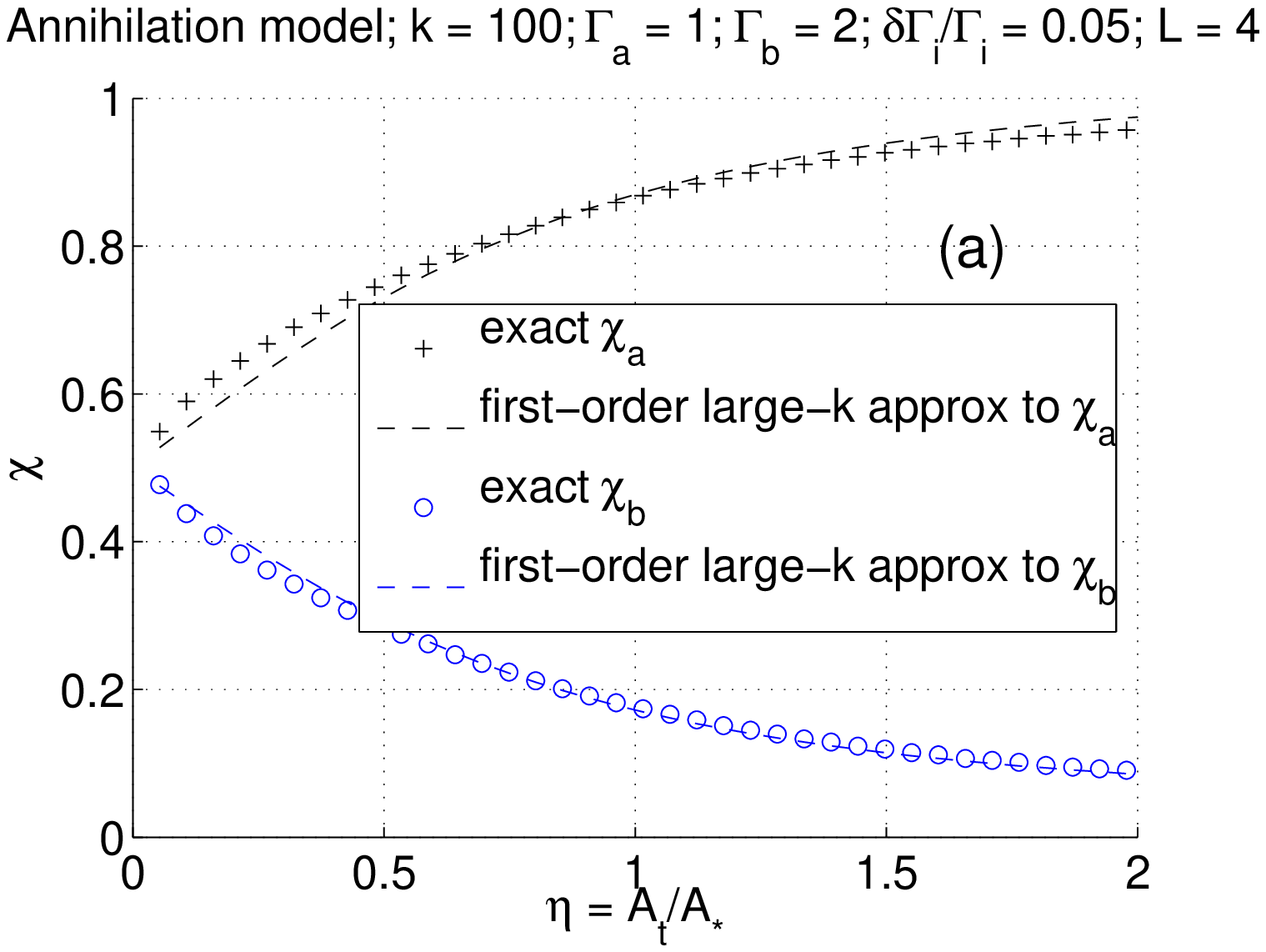}
\\
\includegraphics[scale=0.8]{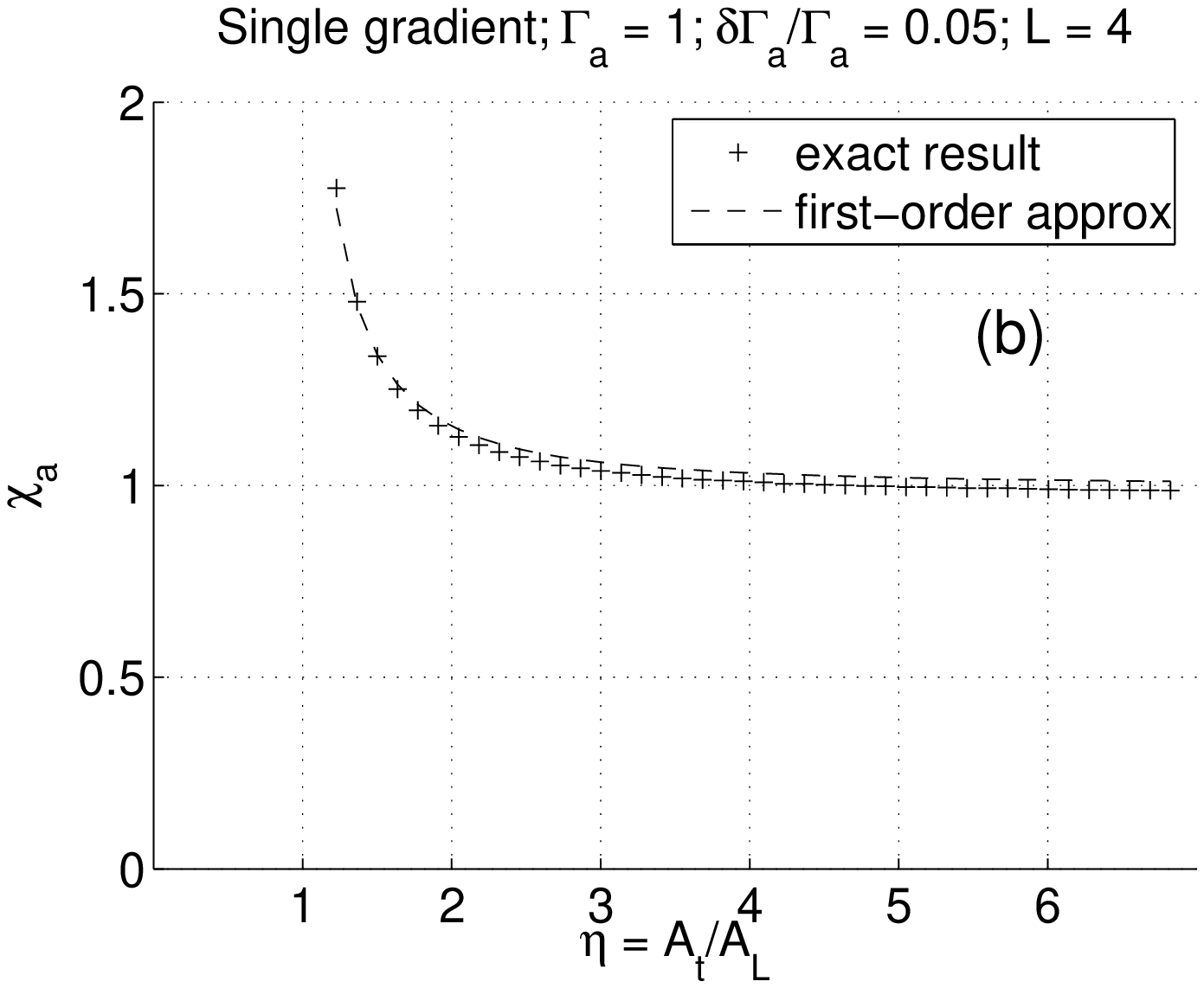}
\end{tabular}
\end{center}
\caption{
Sensitivity of the threshold position $x_t$ to infinitesimal 
variations in the source fluxes $\Gamma_a$ and $\Gamma_b$ in (a) the 
annihilation model (constant diffusion constant and linear degradation) and (b) the single-gradient model. The coefficient of variation $\chi_i$ is
defined by Eq.~(\ref{coeffVar}) in the text. The data plotted as plus signs and circles were obtained 
by solving numerically the full model, while the dashed lines 
represent (from top to bottom) Eqs.~(\ref{chiAEq}),~(\ref{chiBEq}) and~(\ref{chiA_single_eq}). 
All parameters are unity unless otherwise stated. 
\label{chiEta}
}
\end{figure}

The focus of our work has been  the scaling issue. However, we should not lose track of the other requirement for developmental dynamics, namely that the system be relatively robust to fluctuations in parameters such as source fluxes. Fig.~\ref{chiEta}(a) presents data regarding the variation of $x_t$ with $\Gamma_a$ and $\Gamma_b$ in the annihilation model. 
For simplicity the data is
presented for the case of equal decay 
lengths, $\lambda_a = \lambda_b = \lambda$. 
The basic conclusion is that the coefficient of variation $\chi_i$, defined as
\begin{equation}
\frac{\delta x_t}{\lambda} =
\left\{
\begin{array}{l}
 \chi_a \frac{\delta \Gamma_a}{\Gamma_a} \\
- \chi_b \frac{\delta \Gamma_b}{\Gamma_b}
\end{array}
\right.
\label{coeffVar} ,
\end{equation}
starts at $1/2$ at $A_t=0$  and then asymptotes to either $1$ for variations in $\Gamma_a$ or zero for variations in $\Gamma_b$. These asymptotic values are of course precisely the results obtained for the one-exponential-gradient model. The fact that the $\chi_i$ at small $x_t$ is $1/2$ can be understood by noting that in this limit $x_t$ is just $x_e$, which can easily be shown to be approximately (i.e. for large enough $L$) given by $x_e \approx 0.5(L\pm L_c)$ with $L_c = \lambda \left|\ln \left(\Gamma_b/\Gamma_a \right)\right| $. With this approximation for $x_e$ and taking differentials of $x_t$ we obtain 
\beqa
\chi_a & = & \frac{1}{2} + \frac{\eta}{\sqrt{1+\eta^2}} \left[ 1 - \frac{1}{2}\tanh\left(\frac{x_e}{\lambda}\right) \right],\label{chiAEq} \\
\chi_b & = & \frac{1}{2} \left[ 1 - \frac{\eta}{\sqrt{1+\eta^2}}\tanh\left(\frac{x_e}{\lambda}\right) \right] \label{chiBEq}
\eeqa 
where, as before, $\eta = A_t / A_*$. These are good approximations 
at all values of $\eta$ for percentage variations in source fluxes as large as $5\%$ (see Fig.~\ref{chiEta}(a)). 
The reduction of the $\chi$ values  from unity represent an increase in system robustness as compared with the single-exponential-gradient model, albeit with a new sensitivity to the $B$ gradient.
For comparison we also show in Fig.~\ref{chiEta}(b) the coefficient of 
variation which arises in the single-gradient model. The approximation to 
$\chi_a$ in this case is given by
\beq
\chi_a = \frac{\eta}{\sqrt{\eta^2-1}}
\label{chiA_single_eq}
\eeq
where now $\eta$ is defined by $\eta = A_t/A(L)$. Notice that the effect of 
the boundary ($\eta \downarrow 1$) is to increase the sensitivity of the 
gradient to variations in the source flux over that for a simple
exponential.

\section{Conclusions}

In this paper we have shown that coupling two oppositely directed morphogen gradients allows patterns to be set in approximate proportion to the size of the  developing field. We have considered two coupling mechanisms, the most effective of which couples the gradients via a phenomenological annihilation reaction. Such a mechanism can set boundaries of gene expression across the developing field with a small sample-to-sample variation in the normalised position of the boundaries. In this scenario, there is no magic bullet which ensures either exact scaling or complete robustness. Instead, the effective boundary condition created by the annihilation reaction allows for approximate scale invariance to emerge in one reasonably-sized range of parameter space and similarly lowers the sensitivity of any threshold to source-level fluctuations. Presumably, one could obtain even more robustness and scaling, and possibly even temperature compensation (see for example Ref.~\cite{Ismagilov2005}), via the introduction of yet additional interactions. 

After completion of this work we became aware of similar work
in which the annihilation model was applied to 
pattern scaling in the early {\sl Drosophila} embryo 
\cite{tenWoldePrivComm, Howard2005}. 
In contrast to the numerical analysis carried out by the authors of 
Ref.~\cite{Howard2005} for the specific case of the {\sl bicoid} morphogen,
we have presented here a more general 
analytic framework which allows for a natural explanation of 
pattern scaling at intermediate developing-field sizes and 
of filtration of source-level fluctuations. In particular our work provides an explanation for the fact that they found pattern scaling at $L$ approximately 4-5 times the decay length $\lambda$. We note that recent work showing that only {\sl bicoid} binding sites are needed for scaling provides further support for an annihilation mechanism in the {\sl bicoid-hunchback} problem \cite{Dostatni2005}.

\ack
This work has been supported in part by the NSF-sponsored Center
for Theoretical Biological Physics (grant numbers PHY-0216576
and PHY-0225630). 
PM acknowledges useful discussions with E. Levine, T. Hwa and A. Eldar.

\section*{References}

\bibliographystyle{unsrt}
\bibliography{patternScaling}

\newpage 

\end{document}